\documentclass[12pt,letterpaper,superscriptaddress]{revtex4}%
\usepackage{amsmath}
\usepackage{graphicx}
\usepackage{amsfonts}
\usepackage{amssymb}%
\providecommand{\U}[1]{\protect\rule{.1in}{.1in}}
\begin{document}
\title{Spin 1/2 systems perturbed by fluctuating, arbitrary fields; relaxation and
frequency shifts, a new approach to Redfield theory}
\author{R. Golub }
\affiliation{Physics Department, North Carolina State University, Raleigh, NC 27695-8202, USA}
\author{A. Steyerl}
\affiliation{Dept. of Physsics, University of Rhode Island, Kingston. RI 02881, USA}
\date{03/04/14}

\begin{abstract}
The usual approach to considerations of apin relaxation and frequency shifts
due to fluctuating fields is through the density matrix \cite{Slichter}. Here
we treat the problem of the influence of fluctuating fields on a spin 1/2
system based on direct solution of the Schroedinger equation in contrast to
the usual treatment. Our results are seen to be in agreement with the known
results in the literature (\cite{McGregor}, \cite{Slichter}, \cite{Red2},
\cite{CSH}), as they must, but our derivation directly from the Schroedinger
equation allows \ us to see the role of the necessary assumptions in a
somewhat clearer way.

\end{abstract}
\maketitle
\tableofcontents

\section{Introduction \ }

The behavior of a system of spins interacting with static and time varying
magnetic fields is a very broad topic and has been the subject of intense
study for decades. A very important application is to the study of spins
interacting with the randomly fluctuating fields associated with a thermal
reservoir. Bloembergen. Purcell and Pound, \cite{bpp}, have treated this
problem using physical arguments based on Fermi's golden rule and showed that
the relaxation induced by the fields associated with a thermal reservoir is
proportional to the power spectrum of the fluctuating fields evaluated at the
Larmor \ frequency, which is given by the Fourier transform of the
auto-correlation function of these fields. Wangsness and Bloch, \cite{wb}, and
then Bloch, \cite{Bloch}, have approached the problem using second order
perturbation theory applied to the equation of motion of the density matrix
and Redfield, \cite{Red1}, \cite{Red2} (see also \cite{Slichter}) has carried
this calculation forward to show that the relaxation, indeed, depends on the
spectrum of the auto-correlation of the fluctuating fields.

Another source of randomly fluctuating fields is the stochastic motion of
spins (e.g. diffusion) through a region with an inhomogeneous magnetic field.
To study this problem Torrey, \cite{torrey} introduced a diffusion term into
the Bloch equation applied to the bulk magnetization of a sample containing
many spins (Torrey equation). Cates, Schaeffer and Hopper, \cite{CSH} then
rewrote the Torrey equation to apply to the density matrix and solved this
equation to second order in the varying fields using an expansion in the
eigenfunctions of the diffusion equation. McGregor, \cite{McGregor} applied
the Redfield theory to this problem using diffusion theory to calculate the
auto-correlation function of the fluctuating fields seen by spins diffusing
through a (constant gradient) inhomogeneous field. Recently Golub et al,
\cite{Ryan} have shown that these two approaches, \cite{CSH}, \cite{McGregor}
are identical.

A useful review of the field is \cite{Review}.

Another problem which can be treated by these methods is the case of a gas of
spins contained in a vessel subject to inhomogeneous magnetic fields and a
strong electric field as is the case in experiments to search for a non-zero
electric dipole moment of neutral particles such as the neutron,
\cite{LamGolRev} or various atoms or molecules, \cite{AtomMol}. This was shown
by Pendlebury et. al., \cite{JMP PR}, using a second order perturbation
approach to the classical Bloch equation, to lead to an unwanted, linear in
electric field, frequency shift, (often called a 'geometric phase' effect)
which can be the largest systematic error in such experiments

Lamoreaux and Golub, \cite{LamGol} have shown, using a standard density matrix
calculation (Redfield theory), that the 'geometric phase' frequency shift is
given, to second order, by certain correlation functions of the fields seen by
the moving particles.

Pignol and Roccia, \cite{Pignol} have given general results for this effect
valid in the non-adiabatic limit.

Barabanov et al \cite{barab} have given analytic expressions for the relevant
correlation functions for a gas of particles moving in a cylindrical vessel
exposed to a magnetic field with a linear gradient along with an electric
field. Petukhov, et al \cite{Petuk} and Clayton \cite{Clayton} have shown how
to determine the correlation functions for arbitrary geometries and spatial
field dependence for cases where the diffusion theory applies, while Swank et
al, \cite{Swank} have shown how to calculate the spectra of the relevant
correlation functions for gases in rectangular vessels in magnetic fields of
arbitrary position dependence even in those cases where the diffusion theory
does not apply.

Recently Steyerl et al, \cite{Steyerl} have approached the problem of a gas of
spin 1/2 particles subject to time varying magnetic fields by directly solving
the Schroedinger equation to second order. They showed that this approach
leads to the same results as previous work \cite{JMP PR}, \cite{LamGol} for
the 'geometric phase' effect in cylindrical vessels and applied the technique
to several problems of interest such as the frequency shift produced by the
field of a magnetic dipole in the vessel. They have also given solutions for a
general linear gradient as has been discussed in \cite{Pignol}, and higher
order gradients as well.

In the present work we use the methods of \cite{Steyerl} to obtain a general
solution for spin 1/2 valid in all cases where second order perturbation
theory can be applied, including coherent and stochastic fields and long and
short times. In doing this we clarify the meaning of the assumptions necessary
to obtain the Redfield theory.

\section{Solution of the Schroedinger equation for an arbitrary perturbation}

We apply the method introduced by Steyerl et al. starting with the Hamiltonian%
\[
H=-\frac{1}{2}\left[
\begin{array}
[c]{cc}%
\omega_{o}^{\prime} & \omega_{x}-i\omega_{y}\\
\omega_{x}+i\omega_{y} & -\omega_{o}^{\prime}%
\end{array}
\right]  =-\left[
\begin{array}
[c]{cc}%
\omega_{o} & \Omega^{\ast}\\
\Omega & -\omega_{o}%
\end{array}
\right]
\]

where $\omega_{o}^{\prime}=\gamma B_{o},$ $\gamma$ is the gyromagnetic ratio
and $B_{o}$ represents the magnitude of the volume average field in the cell
and the $z$ axis is its direction, $\omega_{o}=\omega_{o}^{\prime}/2,$
$\Omega=\left(  \omega_{x}+i\omega_{y}\right)  /2.$ The Schroedinger equation
is then:%
\begin{equation}
i\frac{\partial}{\partial t}\left[
\begin{array}
[c]{c}%
\alpha\\
\beta
\end{array}
\right]  =H\left[
\begin{array}
[c]{c}%
\alpha\\
\beta
\end{array}
\right]
\end{equation}

Introducing the rotating frame%
\begin{align}
\alpha &  =\alpha_{r}e^{i\omega_{o}\tau}\nonumber\\
\beta &  =\beta_{r}e^{-i\omega_{o}\tau} \label{rot}%
\end{align}%
\begin{align}
-i\dot{\alpha}  &  =\omega_{o}\alpha+\Omega^{\ast}\beta\\
i\dot{\alpha}_{r}  &  =-\Omega^{\ast}\beta_{r}e^{-i2\omega_{o}\tau} \label{2a}%
\end{align}%
\begin{align}
i\dot{\beta}  &  =-\Omega\alpha+\omega_{o}\beta\\
i\dot{\beta}_{r}  &  =-\Omega\alpha_{r}e^{i2\omega_{o}\tau}%
\end{align}%
\begin{align}
i\ddot{\alpha}_{r}  &  =-\dot{\Omega}^{\ast}\beta_{r}e^{-i2\omega_{o}\tau
}-\Omega^{\ast}\dot{\beta}_{r}e^{-i2\omega_{o}\tau}+i2\omega_{o}\Omega^{\ast
}\beta_{r}e^{-i2\omega_{o}\tau}\\
\ddot{\alpha}_{r}-\left(  \frac{\dot{\Omega}^{\ast}}{\Omega^{\ast}}%
-i2\omega_{o}\right)  \dot{\alpha}_{r}  &  =-\left|  \Omega\right|  ^{2}%
\alpha_{r} \label{1}%
\end{align}

\subsection{Perturbation theory}

We now treat the rhs of (\ref{1}) as a perturbation and obtain the zero order
solution by placing this equal to zero:

let%
\begin{equation}
\dot{\alpha}_{r}^{\left(  o\right)  }=y_{o}%
\end{equation}
then%
\begin{align}
\frac{\dot{y}}{y}  &  =\left(  \frac{\dot{\Omega}^{\ast}}{\Omega^{\ast}%
}-i2\omega_{o}\right)  =\frac{d}{dt}\ln y\\
\alpha_{r}^{\left(  0\right)  }  &  =-C_{2}^{\left(  0\right)  }\int
\Omega^{\ast}e^{-i2\omega_{o}t}dt+C_{1}^{\left(  0\right)  }%
\end{align}
Now we substitute this into the rhs of (\ref{1}) to get the next lowest order
solution%
\begin{equation}
\dot{y}_{1}-\left(  \frac{\dot{\Omega}^{\ast}}{\Omega^{\ast}}-i2\omega
_{o}\right)  y_{1}=-\left|  \Omega\right|  ^{2}\left(  -C_{2}^{\left(
0\right)  }\int\Omega^{\ast}e^{-i2\omega_{o}t}dt+C_{1}^{\left(  0\right)
}\right)
\end{equation}
this is of the form%
\begin{equation}
\dot{y}_{1}-P\left(  t\right)  y=q\left(  t\right)
\end{equation}
with%
\begin{align}
P\left(  t\right)   &  =\left(  \frac{\dot{\Omega}^{\ast}}{\Omega^{\ast}%
}-i2\omega_{o}\right) \label{16}\\
q\left(  t\right)   &  =-\left|  \Omega\right|  ^{2}\left(  -C_{2}^{\left(
0\right)  }\int\Omega^{\ast}e^{-i2\omega_{o}t}dt+C_{1}^{\left(  0\right)
}\right)
\end{align}

then by substituting%
\begin{equation}
y=e^{\int Pdt}f
\end{equation}
we find%
\begin{align}
\dot{f}  &  =e^{-\int Pdt}q\left(  t\right) \\
y  &  =e^{\int^{t}Pdt^{\prime}}\int^{t}dt^{\prime}e^{-\int^{t^{\prime}%
}Pdt^{\prime\prime}}q\left(  t^{\prime}\right)
\end{align}
Now, using (\ref{16})%
\begin{align}
\int Pdt  &  =\int dt\left(  \frac{\dot{\Omega}^{\ast}}{\Omega^{\ast}%
}-i2\omega_{o}\right)  =\ln\left(  -\Omega^{\ast}\right)  -i2\omega
_{o}t+K^{\prime}\\
e^{\int^{t}Pdt\prime}  &  =-K\Omega^{\ast}e^{-i2\omega_{o}t}%
\end{align}
and then%
\begin{equation}
y_{1}=-\Omega^{\ast}e^{-i2\omega_{o}t}\int^{t}dt^{\prime}e^{i2\omega
_{o}t^{\prime}}\left[  \Omega\left(  -C_{2}^{\left(  0\right)  }%
\int^{t^{\prime}}\Omega^{\ast}e^{-i2\omega_{o}t^{\prime\prime}}dt^{\prime
\prime}+C_{1}^{\left(  0\right)  }\right)  \right]
\end{equation}

(note $K$ drops out)$.$ The $C_{2}^{\left(  0\right)  }$ term is higher order
in the perturbation so we only have to consider the $C_{1}$ term%
\begin{align}
y_{1}  &  =-C_{1}^{\left(  0\right)  }\Omega^{\ast}e^{-i2\omega_{o}t}\int
^{t}dt^{\prime}e^{i2\omega_{o}t^{\prime}}\left[  \Omega\right] \\
\alpha_{r}^{\left(  1\right)  }  &  =-C_{1}^{\left(  0\right)  }\int
^{t}dt^{\prime}\Omega^{\ast}e^{-i2\omega_{o}t^{\prime}}\int^{t^{\prime}%
}dt^{\prime\prime}e^{i2\omega_{o}t^{\prime\prime}}\left[  \Omega\right]
\end{align}
Combining the two terms for $\alpha_{r}$:%
\begin{equation}
\alpha_{r}=-C_{1}^{\left(  0\right)  }\int^{t}dt^{\prime}\Omega^{\ast}\left\{
e^{-i2\omega_{o}t^{\prime}}\int^{t^{\prime}}dt^{\prime\prime}e^{i2\omega
_{o}t^{\prime\prime}}\left[  \Omega\right]  \right\}  -C_{2}\int^{t}%
\Omega^{\ast}e^{-i2\omega_{o}t^{\prime}}dt^{\prime}+C_{1}^{\left(  0\right)  }%
\end{equation}
and we calculate $\beta_{r}$ from (\ref{2a}):%
\begin{align}
\beta_{r}  &  =-\frac{i\dot{\alpha}_{r}}{\Omega^{\ast}}e^{i2\omega_{o}\tau}\\
&  =i\left(  C_{1}^{\left(  0\right)  }\Omega_{i}\left(  t\right)
+C_{2}\right)
\end{align}
where
\begin{equation}
\Omega_{i}\left(  t\right)  =\int^{t}dt^{\prime}e^{i2\omega_{o}t^{\prime}%
}\Omega\left(  t^{\prime}\right)
\end{equation}

Applying the initial conditions, $\alpha_{r}\left(  0\right)  =1,\beta
_{r}\left(  0\right)  =0$ we have%
\begin{align}
C_{2}  &  =-C_{1}^{\left(  0\right)  }\Omega_{i}\left(  t=0\right) \\
1-C_{1}^{\left(  0\right)  }  &  =C_{1}^{\left(  0\right)  }\left[  \left|
\Omega_{i}\left(  t=0\right)  \right|  ^{2}-F\left(  0\right)  \right]
\end{align}
with%
\begin{align}
F\left(  t\right)   &  =\int^{t}dt^{\prime}\Omega^{\ast}\left(  t^{\prime
}\right)  e^{-i2\omega_{o}t^{\prime}}\Omega_{i}\left(  t^{\prime}\right) \\
\left\langle F\left(  t\right)  -F\left(  t_{o}\right)  \right\rangle  &
=\int_{t_{o}}^{t}dt^{\prime}\int^{t^{\prime}}dt^{\prime\prime}e^{-i2\omega
_{o}\left(  t^{\prime}-t^{\prime\prime}\right)  }\Omega^{\ast}\left(
t^{\prime}\right)  \Omega\left(  t^{\prime\prime}\right)
\end{align}

Then (correct to second order)%
\begin{align}
C_{1}^{\left(  0\right)  }  &  =1-\left|  \Omega_{i}\left(  t=0\right)
\right|  ^{2}+F\left(  0\right) \\
C_{2}  &  =-\Omega_{i}\left(  t=0\right)  \left(  1-\left|  \Omega_{i}\left(
t=0\right)  \right|  ^{2}+F\left(  0\right)  \right)
\end{align}
Putting it together%
\begin{align}
\alpha_{r}  &  =1-\left(  F\left(  t\right)  -F\left(  0\right)  \right)
+\Omega_{i}\left(  0\right)  \left(  \Omega_{i}^{\ast}\left(  t\right)
-\Omega_{i}^{\ast}\left(  0\right)  \right) \nonumber\\
\alpha_{r}  &  =1-\left(  \int_{0}^{t}dt^{\prime}\int_{0}^{t^{\prime}%
}dt^{\prime\prime}e^{-i2\omega_{o}\left(  t^{\prime}-t^{\prime\prime}\right)
}\left(  \Omega^{\ast}\left(  t^{\prime}\right)  \Omega\left(  t^{\prime
\prime}\right)  \right)  \right) \label{0}\\
\beta_{r}  &  =i\left(  \Omega_{i}\left(  t\right)  -\Omega_{i}\left(
0\right)  \right)  =i\int_{0}^{t}dt^{\prime}e^{i2\omega_{o}t^{\prime}}%
\Omega\left(  t^{\prime}\right)  \label{00}%
\end{align}
putting $t_{o}=0$.

The above solution is for a system that starts in the spin up state $\left(
\alpha_{r}\left(  0\right)  =1\right)  .$ Combining with the solution where
the system starts in the spin down state $\left(  \beta_{r}\left(  0\right)
=1\right)  $ we get the general solution in terms of a matrix%

\begin{equation}
\psi_{r}\left(  t\right)  =\left[
\begin{array}
[c]{c}%
a_{r}\left(  t\right) \\
b_{r}\left(  t\right)
\end{array}
\right]  =\left[
\begin{array}
[c]{cc}%
\alpha_{r}\left(  t\right)  & -\beta_{r}^{\ast}\left(  t\right) \\
\beta_{r}\left(  t\right)  & \alpha_{r}^{\ast}\left(  t\right)
\end{array}
\right]  \left[
\begin{array}
[c]{c}%
a\left(  0\right) \\
b\left(  0\right)
\end{array}
\right]  \label{5}%
\end{equation}
where \ the matrix is seen to be unitary if $\alpha_{r},\beta_{r}$ are normalized.

Transforming back to the lab system:%
\begin{equation}
\psi\left(  t\right)  =\left[
\begin{array}
[c]{c}%
e^{i\omega_{o}\tau}a_{r}\left(  t\right) \\
e^{-i\omega_{o}\tau}b_{r}\left(  t\right)
\end{array}
\right]  =\left[
\begin{array}
[c]{cc}%
e^{i\omega_{o}\tau} & 0\\
0 & e^{-i\omega_{o}\tau}%
\end{array}
\right]  \left[
\begin{array}
[c]{cc}%
\alpha_{r}\left(  t\right)  & -\beta_{r}^{\ast}\left(  t\right) \\
\beta_{r}\left(  t\right)  & \alpha_{r}^{\ast}\left(  t\right)
\end{array}
\right]  \left[
\begin{array}
[c]{c}%
a\left(  0\right) \\
b\left(  0\right)
\end{array}
\right]  \label{5a}%
\end{equation}
Equation (\ref{5}) or (\ref{5a}) together with (\ref{0}) and (\ref{00})
represent the complete general solution valid for coherent and incoherent
fluctuating fields and all times, as long as the second order perturbation
approximation is valid, i.e. those times for which the deviations from the
initial values are small (however see below).

\subsection{Example, solution for a constant magnetic field gradient and
constant Electric field ('geometric phase')}

This case is interesting because it results in a serious systematic error in
searches for a particle electric dipole moment \cite{JMP PR}, \cite{LamGol},
\cite{Pignol}, \cite{barab}. In this case%

\begin{equation}
\Omega=a+ibt
\end{equation}
where $a=\frac{\gamma}{2}\left(  \frac{\partial B_{z}}{\partial z}x+\frac
{E}{c}v_{y}\right)  $, $b=\frac{\gamma}{2}\frac{\partial B_{z}}{\partial z}y$,
and the coordinate system is defined so that the particle is moving in the $y$ direction.

Substituting this into equations (\ref{0}) and (\ref{00}) we obtain the
solutions%
\begin{align}
\alpha_{r}  &  =1-z\\
z  &  =\frac{1}{2\omega_{o}^{2}}b^{2}t^{2}+i\left(  -\frac{1}{3\omega_{o}%
}b^{2}t^{3}+\frac{a}{\omega_{o}^{2}}\left(  -\omega_{o}a+b\right)  t\right)
-\frac{1}{\omega_{o}^{4}}\left(  e^{-i\omega_{o}t}\left(  i\left(
b-a\omega_{o}\right)  bt\omega_{o}\right)  \right) \\
\beta_{r}  &  =\left(  \frac{1}{\omega_{o}^{2}}e^{i\omega_{o}t}\left(
-ibt\omega_{o}+\left(  b-a\omega_{o}\right)  \right)  -\frac{1}{\omega_{o}%
^{2}}\left(  b-a\omega_{o}\right)  \right)
\end{align}

This solution is what was obtained in \cite{Steyerl} by a similar method and
was shown there to lead to the known result \cite{JMP PR}, \cite{LamGol} for
the frequency shift. We see that our method (\ref{0}), (\ref{00}) applies to
all times for which the perturbation theory holds, i.e. those times for which
the deviations from the initial values are small.

\section{Phase shifts, frequency shifts and relaxation}

We now consider an ensemble of particles moving on a stochastic set of
trajectories. Each trajectory will be characterized by a given $\Omega\left(
t\right)  $ and we have to take an ensemble average of the frequency shifts
and relaxation rates calculated for each trajectory.

We start by calculating $\sigma_{+}=\left(  \sigma_{x}+i\sigma_{y}\right)  $
and take the initial state to be $\frac{1}{\sqrt{2}}\left(
\begin{array}
[c]{c}%
1\\
1
\end{array}
\right)  $ corresponding to the experimentally common situation of a system
immediately after being exposed to a $\pi/2$ pulse.

so that from (\ref{5})%
\begin{equation}
\psi\left(  t\right)  =\left[
\begin{array}
[c]{c}%
a\left(  t\right) \\
b\left(  t\right)
\end{array}
\right]  =\left[
\begin{array}
[c]{cc}%
\alpha_{r}\left(  t\right)  & -\beta_{r}^{\ast}\left(  t\right) \\
\beta_{r}\left(  t\right)  & \alpha_{r}^{\ast}\left(  t\right)
\end{array}
\right]  \left(
\begin{array}
[c]{c}%
1\\
1
\end{array}
\right)  \frac{1}{\sqrt{2}}=\left[
\begin{array}
[c]{c}%
\alpha_{r}-\beta_{r}^{\ast}\\
\beta_{r}+\alpha_{r}^{\ast}%
\end{array}
\right]  \frac{1}{\sqrt{2}} \label{a}%
\end{equation}

Referring to the wave function (\ref{a}) we evaluate $\sigma_{+}$ in the
rotating frame.
\begin{align}
\left\langle \sigma_{+}\right\rangle _{r}  &  =2a^{\ast}b=\left(  \alpha
_{r}^{\ast}-\beta_{r}\right)  \left(  \beta_{r}+\alpha_{r}^{\ast}\right) \\
&  =\left(  \alpha_{r}^{\ast2}-\beta_{r}^{2}\right)  \label{22}%
\end{align}
Using (\ref{0}, \ref{00})%
\begin{align}
\left\langle \sigma_{+}\right\rangle  &  =1-2\left(  \int_{0}^{t}dt^{\prime
}\int_{0}^{t^{\prime}}dt^{\prime\prime}e^{i2\omega_{o}\left(  t^{\prime
}-t^{\prime\prime}\right)  }\left\langle \Omega\left(  t^{\prime}\right)
\Omega^{\ast}\left(  t^{\prime\prime}\right)  \right\rangle \right)
\nonumber\\
&  +\int_{0}^{t}dt^{\prime}e^{i2\omega_{o}t^{\prime}}\Omega\left(  t^{\prime
}\right)  \int_{0}^{t}dt^{\prime\prime}e^{i2\omega_{o}t^{\prime\prime}}%
\Omega\left(  t^{\prime\prime}\right) \\
&  =1-2\int_{t_{o}}^{t}dt^{\prime}\int_{0}^{t^{\prime}}d\tau e^{i2\omega
_{o}\tau}\left\langle \Omega^{\ast}\left(  t^{\prime}-\tau\right)
\Omega\left(  t^{\prime}\right)  \right\rangle \nonumber\\
&  +\int_{0}^{t}dt^{\prime}e^{i2\omega_{o}\left(  2t^{\prime}\right)  }%
\int_{0}^{t}d\tau e^{-i2\omega_{o}\left(  \tau\right)  }\Omega\left(
t^{\prime}\right)  \Omega\left(  t^{\prime}-\tau\right) \label{33}\\
\left\langle \sigma_{+}\right\rangle  &  =1-2\int_{t_{o}}^{t}dt^{\prime}%
\int_{0}^{t^{\prime}}d\tau e^{i2\omega_{o}\tau}\left\langle \Omega^{\ast
}\left(  t^{\prime}-\tau\right)  \Omega\left(  t^{\prime}\right)
\right\rangle \label{111}%
\end{align}

where the last term in (\ref{33}) vanishes because the integrand is a rapidly
varying function of $t^{\prime}.$ From the behavior of $\left\langle
\sigma_{+}\right\rangle $ we can obtain the frequency shift, $\delta\omega,$
and the transverse relaxation rate, $1/T_{2}$.

\subsection{Phase shifts and frequency shifts}

Now $\left\langle \sigma_{+}\right\rangle =1+z_{2}$ where $z_{2}=$
$z_{2}^{\prime}+iz_{2}^{\prime\prime}$ is second order in the perturbation, so
that (from (\ref{111}))
\begin{align}
\left\langle \delta\phi\right\rangle  &  =\arg\left\langle \sigma
_{+}\right\rangle =\arg\left(  1+z_{2}\right)  =\tan^{-1}\left(  \frac
{z_{2}^{\prime\prime}}{1+z_{2}^{\prime}}\right) \nonumber\\
&  \simeq z_{2}^{\prime\prime}=-2\operatorname{Im}\int_{t_{o}}^{t}dt^{\prime
}\int_{0}^{t^{\prime}}d\tau e^{i2\omega_{o}\tau}\left\langle \Omega^{\ast
}\left(  t^{\prime}-\tau\right)  \Omega\left(  t^{\prime}\right)
\right\rangle \\
&  =2\operatorname{Im}\int_{t_{o}}^{t}dt^{\prime}\int_{0}^{t^{\prime}}d\tau
e^{-i2\omega_{o}\tau}\left\langle \Omega\left(  t^{\prime}-\tau\right)
\Omega^{\ast}\left(  t^{\prime}\right)  \right\rangle
\end{align}

Then differentiating w.r.t. $t$ to get the frequency shift we have
\begin{equation}
\delta\omega=2\operatorname{Im}\left(  \int_{0}^{t}d\tau e^{-i2\omega_{o}\tau
}\left\langle \Omega^{\ast}\left(  t\right)  \Omega\left(  t-\tau\right)
\right\rangle \right)
\end{equation}

\begin{align}
\delta\omega &  =\frac{1}{2}\operatorname{Im}\left(  \int_{0}^{t}d\tau\left(
\begin{array}
[c]{c}%
\cos\omega_{o}^{\prime}\tau\\
-i\sin\omega_{o}^{\prime}\tau
\end{array}
\right)  \left\langle \left(  \omega_{x}\left(  t\right)  -i\omega_{y}\left(
t\right)  \right)  \left(  \omega_{x}\left(  t-\tau\right)  +i\omega
_{y}\left(  t-\tau\right)  \right)  \right\rangle \right) \nonumber\\
&  =\frac{1}{2}\int_{0}^{t}d\tau\left(
\begin{array}
[c]{c}%
\cos\omega_{o}^{\prime}\tau\left\langle \omega_{x}\left(  t\right)  \omega
_{y}\left(  t-\tau\right)  -\omega_{y}\left(  t\right)  \omega_{x}\left(
t-\tau\right)  \right\rangle +\\
-\sin\omega_{o}^{\prime}\tau\left\langle \omega_{x}\left(  t\right)
\omega_{x}\left(  t-\tau\right)  +\omega_{y}\left(  t\right)  \omega
_{y}\left(  t-\tau\right)  \right\rangle
\end{array}
\right)  \label{8}%
\end{align}
which is in agreement with previous results, \cite{JMP PR}, \cite{LamGol},
\cite{barab}, \cite{Pignol}.

There has been some discussion in the literature, \cite{Pignol}, concerning
the correct signs in this expression. After discussions with the author,
\cite{Pignol2} and reworking of some previous calculations we have shown that
all results agree with (\ref{8}).

\subsubsection{An assumption of Redfield theory}

Redfield and other authors \cite{Red1}. \cite{Slichter} have taken $t$ large
enough in (\ref{111}) so that the correlation functions vanish at that time,
i.e. $t>\tau_{c}$ where $\tau_{c}$ is the time it takes $\left\langle
\Omega^{\ast}\left(  t\right)  \Omega\left(  t-\tau\right)  \right\rangle $ to
go to zero and the upper limit of integration can then be taken to be
infinite. This then results in the the integral giving the Fourier transform
of the correlation function of the fluctuating field as introduced by
Bloembergen, Pound and Purcell. However as is well known (see \cite{Slichter})
this step is not necessary, it is introduced only to allow writing the results
in terms of the Fourier transform, the results (\ref{0} and \ref{00}) are
valid for short times as well and also apply to the case of coherent fields as
shown above.

\subsection{T$_{2}$ Relaxation}

\bigskip With $\left\langle \sigma_{+}\right\rangle =1+z_{2}$ we calculate
\begin{align}
\left|  \left\langle \sigma_{+}\right\rangle \right|  ^{2}  &
=1+2\operatorname{Re}z_{2}=1+z_{2}+z_{2}^{\ast}\\
\left|  \left\langle \sigma_{+}\right\rangle \right|   &  =1+\left(
z_{2}+z_{2}^{\ast}\right)  /2=1-\int_{t_{o}}^{t}dt^{\prime}\int_{t_{o}%
}^{t^{\prime}}dt^{\prime\prime}e^{i2\omega_{o}\left(  t^{\prime}%
-t^{\prime\prime}\right)  }\left\langle \Omega^{\ast}\left(  t^{\prime\prime
}\right)  \Omega\left(  t^{\prime}\right)  \right\rangle -c.c.\nonumber
\end{align}

using (\ref{111}). $c.c.$ is the complex conjugate of the second term.

We now specialize to the case of a stationary system where $\left\langle
\Omega^{\ast}\left(  t^{\prime}\right)  \Omega\left(  t^{\prime\prime}\right)
\right\rangle $ is a function of $\left(  t^{\prime}-t^{\prime\prime}\right)
$ only. Consider a square region of the $t^{\prime\prime},t^{\prime}$ plane
between $\left(  t_{o},t_{o}\right)  $, $\left(  t_{o},t\right)  ,\left(
t,t_{o}\right)  $ and $\left(  t,t\right)  $. (See (\cite{Squires}) \ for a
discussion of this argument). Then the double integral over the top half
$\left(  t^{\prime}>t^{\prime\prime}\right)  $ is seen to be the complex
conjugate of the integral over the bottom half $\left(  t^{\prime}%
<t^{\prime\prime}\right)  $, so the last two terms are given by the integral
over the entire square

As a result of this we have\bigskip\ (again putting $t^{\prime}-t^{\prime
\prime}=\tau$)%
\begin{align}
\left|  \sigma_{+}\right|   &  =1-\int_{t_{o}}^{t}dt^{\prime}\int_{t_{o}}%
^{t}dt^{\prime\prime}e^{-i2\omega_{o}\left(  t^{\prime}-t^{\prime\prime
}\right)  }\left\langle \Omega^{\ast}\left(  t^{\prime}\right)  \Omega\left(
t^{\prime\prime}\right)  \right\rangle \nonumber\\
&  =1-\int_{-t}^{t}d\tau\left(  t-\left|  \tau\right|  \right)  e^{-i2\omega
_{o}\tau}\left\langle \Omega^{\ast}\left(  t^{\prime}\right)  \Omega\left(
t^{\prime}-\tau\right)  \right\rangle \nonumber\\
&  =1-t\int_{-t}^{t}d\tau e^{-i2\omega_{o}\tau}\left\langle \Omega^{\ast
}\left(  t^{\prime}\right)  \Omega\left(  t^{\prime}-\tau\right)
\right\rangle \label{oooa}\\
\left|  \sigma_{+}\right|   &  =1-\frac{t}{T_{2}} \label{00a}%
\end{align}

\subsubsection{Comparison to Redfield theory}

The step leading to (\ref{oooa}) is based on taking $t\gg\tau_{c}$ (following
Redfield), where $\tau_{c}$ is the correlation time or the time that it takes
$\left\langle \Omega^{\ast}\left(  t^{\prime}\right)  \Omega\left(  t^{\prime
}-\tau\right)  \right\rangle $ to go to zero. (See above ). For shorter times
we would not have a linear but (for, say, the non-adiabatic limit, $\omega
_{o}\tau_{c}<<1$), a quadratic decay

The result (\ref{00a}) obtained in second order perturbation theory is valid
only as long as subsequent terms can be neglected. This requires that $\left(
t/T_{2}<<1\right)  $ or that the changes in the wave function remain small. In
the Redfield treatment we assume that we are dealing with times short enough
that we can replace $\rho(0)$ by $\rho\left(  t\right)  $ in the equation for
$\dot{\rho}(t)$ ($\rho\left(  t\right)  $ is the spin density matrix)
obtaining an equation%
\begin{equation}
\frac{\partial\rho}{\partial t}=\Gamma\cdot\rho\left(  t\right)
\end{equation}

where $\Gamma$ is the 'relaxation matrix'. This equation is then valid for
times so long that the changes in the system are significant as discussed by
Slichter, p. 204 \cite{Slichter}.

In our case we can formulate the argument in a slightly different way.
Consider (\ref{00a}) after a time $\delta t$,
\begin{equation}
\left|  \sigma_{+}\right|  =1-\frac{\delta t}{T_{2}}%
\end{equation}
as the initial condition for the interval $t=\delta t$ to $t=2\delta t$ after
which time we will have%
\begin{equation}
\left|  \sigma_{+}\right|  =\left(  1-\frac{\delta t}{T_{2}}\right)  ^{2}%
\end{equation}
Continuing the argument, after a time $t$ we will have%
\begin{equation}
\left|  \sigma_{+}\right|  =\left(  1-\frac{\delta t}{T_{2}}\right)
^{\frac{t}{\delta t}}=e^{-t/T_{2}}%
\end{equation}
if we take the limit as $\delta t\rightarrow0.$

\subsection{\bigskip T$_{2}$ Relaxation continued}

Thus from (\ref{oooa}) and (\ref{00a})
\begin{align}
\frac{1}{T_{2}}  &  =\int_{-\infty}^{\infty}d\tau e^{-i2\omega_{o}\tau
}\left\langle \Omega^{\ast}\left(  t^{\prime}\right)  \Omega\left(  t^{\prime
}-\tau\right)  \right\rangle \\
&  =\frac{1}{4}\int_{-\infty}^{\infty}d\tau\left(
\begin{array}
[c]{c}%
\cos\omega_{o}\tau\left\langle \omega_{x}\left(  t\right)  \omega_{x}\left(
t-\tau\right)  +\omega_{y}\left(  t\right)  \omega_{y}\left(  t-\tau\right)
\right\rangle \\
+\sin\omega_{o}\tau\left\langle \omega_{y}\left(  t\right)  \omega_{x}\left(
t-\tau\right)  -\omega_{x}\left(  t\right)  \omega_{y}\left(  t-\tau\right)
\right\rangle
\end{array}
\right)  \label{222}%
\end{align}
where the imaginary terms vanish as expected because their integrands are odd.
We have replaced $\omega_{i}$ by $-\omega_{i}/2$ as discussed above. The
second term is absent in the usual treatments as it is normally assumed that
the cross correlation between the components of the fluctuating field vanishes.

\subsubsection{Contribution of fluctuating B$_{z}$}

For simplicity we consider the effects of a fluctuating $B_{z}$ independently
of the other components and will add the results.

In that case the Hamiltonian is:%
\begin{equation}
H=-\frac{1}{2}\left|
\begin{array}
[c]{cc}%
\omega_{o}^{\prime}+\omega_{z}^{\prime} & 0\\
0 & -\left(  \omega_{o}^{\prime}+\omega_{z}^{\prime}\right)
\end{array}
\right|
\end{equation}

Here $\omega_{o}^{\prime}$ represents the average value of $B_{z}$ while
$\omega_{z}^{\prime}$ corresponds to the fluctuations around this average. The
Schroedinger equation is $\left(  \omega_{i}=\omega_{i}^{\prime}/2\right)  $:
\begin{align}
i\left[
\begin{array}
[c]{c}%
\dot{\alpha}\\
\dot{\beta}%
\end{array}
\right]   &  =-\left[
\begin{array}
[c]{c}%
\left(  \omega_{o}+\omega_{z}\right)  \alpha\\
-\left(  \omega_{o}+\omega_{z}\right)  \beta
\end{array}
\right] \\
i\dot{\alpha}  &  =-\left(  \omega_{o}+\omega_{z}\right)  \alpha\\
i\dot{\alpha}_{r}  &  =-\omega_{z}\alpha_{r}%
\end{align}
using (\ref{rot}). Then
\begin{align}
\ln\left(  \alpha_{r}\right)   &  =i\int_{0}^{t}\omega_{z}dt^{\prime}+C\\
\alpha_{r}  &  =e^{i\int_{0}^{t}\omega_{z}dt^{\prime}}\\
\beta_{r}  &  =0
\end{align}
since we want a solution that is in the $\sigma_{z}=+1$ state at $t=0.$

Combining with the solution for $\sigma_{z}=-1$ as the initial state we have,
now taking a state with the spin along the $x$ axis as the initial state%
\[
\psi\left(  t\right)  =\left[
\begin{array}
[c]{cc}%
\alpha_{r} & 0\\
0 & \alpha_{r}^{\ast}%
\end{array}
\right]  \left[
\begin{array}
[c]{c}%
1\\
1
\end{array}
\right]  =\frac{1}{\sqrt{2}}\left[
\begin{array}
[c]{c}%
\alpha_{r}\\
\alpha_{r}^{\ast}%
\end{array}
\right]
\]
so that
\begin{align}
\left\langle \sigma_{+}\right\rangle  &  =\left(  \alpha_{r}^{\ast}\right)
^{2}=\left\langle e^{-2i\int_{0}^{t}\omega_{z}dt^{\prime}}\right\rangle \\
&  \approx1-2i\left\langle \int_{0}^{t}\omega_{z}dt^{\prime}\right\rangle
-2\left\langle \left[  \int_{0}^{t}\omega_{z}dt^{\prime}\right]
^{2}\right\rangle \\
&  \approx1-2\left\langle \int_{0}^{t}\omega_{z}dt^{\prime}\int_{0}^{t}%
\omega_{z}dt^{\prime\prime}\right\rangle
\end{align}
where we used the fact that the average of the fluctuating fields is zero by
definition. Thus%
\begin{align}
\left\langle \sigma_{+}\right\rangle  &  \approx1-2\int_{0}^{t}dt^{\prime}%
\int_{0}^{t}dt^{\prime\prime}\left\langle \omega_{z}\left(  t^{\prime}\right)
\omega_{z}\left(  t^{\prime\prime}\right)  \right\rangle \\
&  \approx1-2\int_{-t}^{t}d\tau\left(  t-\left|  \tau\right|  \right)
\left\langle \omega_{z}\left(  t^{\prime}\right)  \omega_{z}\left(  t^{\prime
}-\tau\right)  \right\rangle \\
&  \approx1-2t\int_{-\infty}^{\infty}d\tau\left\langle \omega_{z}\left(
0\right)  \omega_{z}\left(  \tau\right)  \right\rangle
\end{align}
where we have again taken $\left(  t>\tau_{c}\right)  $ in order to obtain the
Fourier transform. We now have%
\begin{equation}
\frac{1}{T_{2}^{\left(  z\right)  }}=\frac{1}{2}\int_{-\infty}^{\infty}%
d\tau\left\langle \omega_{z}^{\prime}\left(  0\right)  \omega_{z}^{\prime
}\left(  \tau\right)  \right\rangle
\end{equation}
which is to be added to (\ref{222}) to obtain the total transverse relaxation rate.

\section{\bigskip T$_{1}$ Relaxation}

\bigskip To calculate the $T_{1}$ relaxation we start in the up state:%
\begin{equation}
\psi\left(  t\right)  =\left[
\begin{array}
[c]{c}%
a\left(  t\right) \\
b\left(  t\right)
\end{array}
\right]  =\left[
\begin{array}
[c]{cc}%
\alpha_{r}\left(  t\right)  & -\beta_{r}^{\ast}\left(  t\right) \\
\beta_{r}\left(  t\right)  & \alpha_{r}^{\ast}\left(  t\right)
\end{array}
\right]  \left(
\begin{array}
[c]{c}%
1\\
0
\end{array}
\right)  =\left[
\begin{array}
[c]{c}%
\alpha_{r}\\
\beta_{r}%
\end{array}
\right]  \label{aa}%
\end{equation}
and calculate%

\begin{equation}
\left\langle \sigma_{z}\right\rangle =\left|  a\right|  ^{2}-\left|  b\right|
^{2}=\alpha_{r}\alpha_{r}^{\ast}-\beta_{r}\beta_{r}^{\ast}%
\end{equation}
From (\ref{0})%

\begin{align}
\alpha_{r}  &  =1-\left(  \int_{0}^{t}dt^{\prime}\int_{0}^{t^{\prime}%
}dt^{\prime\prime}e^{-i2\omega_{o}\left(  t^{\prime}-t^{\prime\prime}\right)
}\left\langle \Omega^{\ast}\left(  t^{\prime}\right)  \Omega\left(
t^{\prime\prime}\right)  \right\rangle \right) \\
&  =1-\varepsilon_{2}\\
\alpha_{r}\alpha_{r}^{\ast}  &  =1-\left(  \varepsilon_{2}+\varepsilon
_{2}^{\ast}\right)  =1-2\operatorname{Re}\int_{0}^{t}dt^{\prime}\int
_{0}^{t^{\prime}}dt^{\prime\prime}e^{-i2\omega_{o}\left(  t^{\prime}%
-t^{\prime\prime}\right)  }\left\langle \Omega^{\ast}\left(  t^{\prime
}\right)  \Omega\left(  t^{\prime\prime}\right)  \right\rangle \\
&  =1-\int_{0}^{t}dt^{\prime}\int_{0}^{t}dt^{\prime\prime}e^{-i2\omega
_{o}\left(  t^{\prime}-t^{\prime\prime}\right)  }\left\langle \Omega^{\ast
}\left(  t^{\prime}\right)  \Omega\left(  t^{\prime\prime}\right)
\right\rangle
\end{align}
as shown above. From (\ref{00})%

\begin{align}
\beta_{r}  &  =i\int_{0}^{t}dt^{\prime}e^{i2\omega_{o}t^{\prime}}\Omega\left(
t^{\prime}\right) \\
\beta_{r}\beta_{r}^{\ast}  &  =\int_{0}^{t}dt^{\prime}\int_{0}^{t}%
dt^{\prime\prime}\left\langle \Omega^{\ast}\left(  t^{\prime}\right)
\Omega\left(  t^{\prime\prime}\right)  \right\rangle e^{-i2\omega_{o}\left(
t^{\prime}-t^{\prime\prime}\right)  }%
\end{align}

\begin{align}
\left\langle \sigma_{z}\right\rangle  &  =\alpha_{r}\alpha_{r}^{\ast}%
-\beta_{r}\beta_{r}^{\ast}\\
&  =1-2\int_{0}^{t}dt^{\prime}\int_{0}^{t}dt^{\prime\prime}\left\langle
\Omega^{\ast}\left(  t^{\prime}\right)  \Omega\left(  t^{\prime\prime}\right)
\right\rangle e^{i2\omega_{o}\left(  t^{\prime}-t^{\prime\prime}\right)  }%
\end{align}%
\begin{align}
&  =1-2\int_{-t}^{t}d\tau\left(  t-\left|  \tau\right|  \right)
e^{-i2\omega_{o}\tau}\left\langle \Omega^{\ast}\left(  t^{\prime}\right)
\Omega\left(  t^{\prime}-\tau\right)  \right\rangle \\
&  =1-t2\int_{-\infty}^{\infty}d\tau e^{-i2\omega_{o}\tau}\left\langle
\Omega^{\ast}\left(  t^{\prime}\right)  \Omega\left(  t^{\prime}-\tau\right)
\right\rangle
\end{align}
again specializing to $\left(  t>>\tau_{c}\right)  $

\bigskip%
\begin{align}
\frac{1}{T_{1}}  &  =2\int_{-\infty}^{\infty}d\tau e^{-i2\omega_{o}\tau
}\left\langle \Omega^{\ast}\left(  t^{\prime}\right)  \Omega\left(  t^{\prime
}-\tau\right)  \right\rangle \\
&  =\frac{1}{2}\int_{-\infty}^{\infty}d\tau\left(
\begin{array}
[c]{c}%
\cos\omega_{o}^{\prime}\tau\\
-i\sin\omega_{o}^{\prime}\tau
\end{array}
\right)  \left\langle \left(  \omega_{x}\left(  t^{\prime}\right)
-i\omega_{y}\left(  t^{\prime}\right)  \right)  \left(  \omega_{x}\left(
t^{\prime}-\tau\right)  +i\omega_{y}\left(  t^{\prime}-\tau\right)  \right)
\right\rangle \nonumber\\
&  =\frac{1}{2}\int_{-\infty}^{\infty}d\tau\left(
\begin{array}
[c]{c}%
\cos\omega_{o}^{\prime}\tau\left\langle \omega_{x}\left(  t\right)  \omega
_{x}\left(  t-\tau\right)  +\omega_{y}\left(  t\right)  \omega_{y}\left(
t-\tau\right)  \right\rangle \\
+\sin\omega_{o}^{\prime}\tau\left\langle \omega_{y}\left(  t\right)
\omega_{x}\left(  t-\tau\right)  -\omega_{x}\left(  t\right)  \omega
_{y}\left(  t-\tau\right)  \right\rangle
\end{array}
\right) \label{222a}\\
&  =\frac{2}{T_{2}^{\prime}}%
\end{align}
$1/T_{2}^{\prime}$ is the relaxation rate without the contribution of the
fluctuations in $B_{z}$ (\ref{222}).

\section{Conclusion}

We have treated the problem of the influence of fluctuating fields on a spin
1/2 system based on direct solution of the Schroedinger equation in contrast
to the usual treatment based on the density matrix (Redfield theory)

Our results are seen to be in agreement with the known results in the
literature (\cite{McGregor}, \cite{Slichter}, \cite{Red2}, \cite{CSH}), as
they must, but our derivation directly from the Schroedinger equation allows
\ us to see the role of the necessary assumptions in a somewhat clearer way.

To get the Redfield results from the general solution it is necessary to
assume the field fluctuations are stationary and to limit ourselves to times
much longer than the correlation time. However this is only necessary to get
the result in the satisfying form of a Fourier transform. The general solution
will be valid for times shorter than the correlation time as well. The
requirements of second order perturbation theory that the change in the wave
function must remain small can be relaxed by treating changes over consecutive
small time periods similar to what is done in the density matrix treatment,
\cite{Slichter}.

Our results (\ref{0}) and (\ref{00}) are very general and can be applied to
coherent and stochastic fields also in the case of short times.

The density matrix was introduced to simplify the treatment of 'mixed' states,
states described by an ensemble of systems in 'pure' quantum states, i.e.
systems where some parameter, e.g. a phase, is a stochastic variable. However
the same results can always be obtained by calculating the wave function as if
for a pure state and then averaging the results for observables over the
stochastic parameters. In general, the solution of the Schroedinger equation
is easier than the solution of the equation for the density matrix, but the
calculation of observables (expectation values) from the results is easier in
the case of the density matrix. \ Since the most difficult step is usually
solving the differential equations we would argue that the wave function
approach presented here is more often advantageous.


\begin{thebibliography}{99}                                                                                               %


\bibitem {bpp}Bloembergen, N., Pound, E.M, and Purcell, R.V., Phys.
Rev.\textbf{73}, 679 (1948) \ 

\bibitem {wb}Wangsness R.K. and Bloch, F., Phys. Rev \textbf{89}, 728, (1953)

\bibitem {Bloch}Bloch, F. Phys. Rev. \textbf{102, }104, (1956)

\bibitem {Red1}Redfield, A.G., IBM J. Res. Dev. \textbf{1}, 19 (1957)

\bibitem {Red2}Redfield, A.G. Advances in Magnetic Resonance, vol \textbf{1},
1 (1965) Academic Press

\bibitem {Slichter}Slichter, C.P., ''Principles of magnetic resonance'', Third
edition, Springer Verlag, 1990

\bibitem {torrey}Torrey, H.C., Phys. Rev. \textbf{104}, 563 (19560

\bibitem {CSH}Cates, G.D., Schaeffer, S.R. and Happer,\ W., Phys. Rev.
\textbf{A37}, 2877 (1988)

\bibitem {McGregor}McGregor, D.D., Phys. Rev. \textbf{A41}, 2631. (1990)

\bibitem {Ryan}Golub, R., Rohm, R.M. and Swank, C.M., Phys. Rev. \textbf{A83},
023402, (2011)

\bibitem {Review}Nicholas, M.P., Erylmas, E., Ferrage, F., Cowburn, D. and
Ghose, R., Progress in Nuclear Magnetic Resonance Spectroscopy, \textbf{57},
111 (2010)

\bibitem {LamGolRev}Lamoraux, S.K. and Golub, R., J. Phys. G: Nucl. Part.
Phys. \textbf{36,} 104002, (2009)

\bibitem {AtomMol}Eckel, S., Hamilton, P., Kirilov, E., Smith, H. W. and D.
DeMille, D.

Phys. Rev. \textbf{A 87}, 052130, (2013)

\bibitem {JMP PR}Pendlebury, J.M. et al, Phys. Rev. \textbf{A 70}, 032102 (2004)

\bibitem {LamGol}Lamoreaux. S.K. and Golub, R. Phys. Rev. \textbf{A 71},
032104 (2005)

\bibitem {Pignol}Pignol, G. and Roccia, S., Phys. Rev. \textbf{A85, },042105. (2012)

\bibitem {Pignol2}Pignol, G. \emph{Private Communication, Feb. 2014}

\bibitem {barab}Barabanov, A.L., Golub, R. and Lamoreaux, S.K., Phys. Rev.
\textbf{A 74}, 052115 (2006)

\bibitem {Petuk}Petukhov, A.K. Pignol, G. Jullien, D. and Andersen, K.H.,
Phys. Rev. Lett. 105, 170401 (2010)

\bibitem {Clayton}Clayton, S.M., J. of Mag. Res., \textbf{211,} 89 (2011)

\bibitem {Swank}Swank, C.M. Petukhov, A.K. and Golub,R., Phys Letts \textbf{A
376,} 2319 (2012)

\bibitem {Steyerl}Steyerl. A., Kaufman, C., M\"{u}ller, G., Malik, S.S.,
Desai, M. and Golub, R., to be published

\bibitem {Squires}Squires, G.L.. \ \emph{\ \ Introduction to the theory of
thermal neutron scattering, }CUP, 1978, pp 98-99
\end{thebibliography}
\end{document}